\documentclass[preprint2]{aastex}

\shorttitle{The Re50N Eruptive Event}
\shortauthors{Chiang et al.}

\begin{document}

\title{The Brightening of Re50N: \\
Accretion Event or Dust Clearing?}

\author{Hsin-Fang Chiang\altaffilmark{1},
        Bo Reipurth\altaffilmark{1}, 
        Josh Walawender\altaffilmark{2},        
        Michael S. Connelley\altaffilmark{1},\\
        Peter Pessev\altaffilmark{3},
        T.R. Geballe\altaffilmark{4},
        William M.J. Best\altaffilmark{5},
        Martin Paegert\altaffilmark{6}        
        }

\vspace{0.5cm}

\affil{1: Institute for Astronomy, University of Hawaii at Manoa,\\ 
          640 N. Aohoku Place, HI 96720, USA}
  \email{hchiang/reipurth@ifa.hawaii.edu}

\affil{2: Subaru Telescope, National Astronomical Observatory of Japan, Hilo, HI 96720, USA}
  \email{joshw@naoj.org}

\affil{3: Gemini Observatory, Casilla 603, La Serena, Chile}
  \email{pessev@gmail.com}

\affil{4: Gemini Observatory, 670 N. Aohoku Place, Hilo, HI 96720, USA}
  \email{tgeballe@gemini.edu}

\affil{5: Institute for Astronomy, 2680 Woodlawn Drive, Honolulu, HI 96822, USA}
  \email{wbest@ifa.hawaii.edu}

\affil{6: Department of Physics \& Astronomy, Vanderbilt University,
Nashville, TN 37235, USA}
  \email{martin.paegert@vanderbilt.edu}

\newcommand{\simless}{\mathbin{\lower 3pt\hbox
     {$\rlap{\raise 5pt\hbox{$\char'074$}}\mathchar"7218$}}}

\begin{abstract}

  The luminous Class~I protostar HBC~494, embedded in the Orion~A
  cloud, is associated with a pair of reflection nebulae, Re50 and
  Re50N, which appeared sometime between 1955 and 1979. We have found
  that a dramatic brightening of Re50N has taken place sometime
  between 2006 and 2014. This could result if the embedded source is
  undergoing a FUor eruption. However, the near-infrared spectrum
  shows a featureless very red continuum, in contrast to the strong CO
  bandhead absorption displayed by FUors. Such heavy veiling, and the
  high luminosity of the protostar, is indicative of strong accretion
  but seemingly not in the manner of typical FUors. We favor the
  alternative explanation that the major brightening of Re50N and the
  simultaneous fading of Re50 is caused by curtains of obscuring
  material that cast patterns of illumination and shadows across the
  surface of the molecular cloud. This is likely occurring as an
  outflow cavity surrounding the embedded protostar breaks through to
  the surface of the molecular cloud. Several Herbig-Haro objects are
  found in the region.

\end{abstract}


\section{INTRODUCTION}

Stars are born deeply embedded in dense cloud cores, which are
themselves embedded in larger, more tenuous molecular clouds. As a
newborn star gains mass, it forms a circumstellar disk, and
perpendicular to the disk the star drives a powerful bipolar outflow.
The outflow creates cavities in the molecular surroundings, and when a
cavity breaks through the surface of the surrounding cloud, it allows
light to escape from the cloud for the first time. As the
cavities expand, the star eventually emerges from its confinement.
The first manifestation of the opening of an outflow cavity is the
appearance of a compact reflection nebula, as the light of the star
floods the surrounding molecular landscape. Such nebulae are commonly
found when inspecting the surfaces of dark clouds.  

A particularly fine example of a reflection nebula was discovered in
the southern L1641 cloud (Reipurth 1985).  Examination of Schmidt
plates of the region showed that sometime between 1955 and 1979 a
bright $\sim$1~arcmin wide reflection nebula appeared next to the
embedded source IRAS 05380--0728 in the southern part of the Orion~A
cloud complex.  In a more detailed study, Reipurth \& Bally (1986)
obtained an optical red spectrum of the reflection nebula, showing
H$\alpha$ and the Ca triplet in emission. With a luminosity of
$\sim$250~L$_\odot$, the source, labeled HBC~494 by Herbig \& Bell
(1988), is among the most luminous sources in the L1641 cloud. The
nebula has two components, a larger part, Re50, and a highly variable
part to the north, Re50N; the latter is associated with the embedded
source.  A molecular outflow is associated with the source (Reipurth
\& Bally 1986, Fukui et al.  1986), and has been studied in detail by
Lee et al. (2002).

The region has since been explored in a number of studies.  Scarrott
\& Wolstencroft (1988) used optical polarization imaging to identify
the location of the source HBC~494, which was subsequently detected as
a radio continuum source (VLA~1) by Morgan et al. (1990), and a second
radio continuum source VLA~2 was found by Anglada (1995) about an
arcminute to the WNW of VLA~1.  Casali (1991) determined an extinction
of A$_V$$\sim$50 mag to the illuminating source whereas Quanz et al.
(2007) found A$_V$$\sim$26 mag, and Colom\'e et al.  (1996) suggested
a total mass of $\sim$3$\times$10$^3$~M$_\odot$ within 0.3~pc of the
source.

In this paper we document a major recent brightening of Re50N and
discuss the possible underlying mechanisms.

\section{OBSERVATIONS}

The Variable Young Stellar Objects Survey (VYSOS) program employs two
robotic telescopes, of which a 20 inch f/8.2 Ritchey-Chretien
reflector (VYSOS-20) is used for the observations reported here. The
location of the telescope is at Mauna Loa Observatory in Hawaii, at a
latitude of 20$^\circ$ and an elevation of 3400m. Imaging is done with
an Apogee Alta U16M CCD and the field of view is 30.8'$\times$30.8'
with 0.45''/pixel. VYSOS-20 started regular operation in July 2012,
with automatic scheduling software to repeatedly observe selected star
forming regions.  In November 2013, we noticed an increase of the
brightness of Re50N, and started to monitor the target each night.
Three exposures of 100 sec are taken in a Sloan {\em i}-band filter
for each target every night.

The imaging data were calibrated using the Image Reduction and
Analysis Facility (IRAF; Tody 1986).  Specifically, the aperture
photometry was performed using IMEXAMINE in IRAF, and since Re50N is
extended an aperture size of 17.8'' was used to include all the main
bright nebulosity.  The absolute magnitude scale was established based
on reference stars from the Pan-STARRS catalogues (E. Magnier, private
communications).  The photometric uncertainty is estimated from the
Pythagorean sum of the standard errors from each night and an
additional calibration error measured using a nearby star.

Five 2-min exposures were made of the Re50 region with the 8~m Subaru
telescope and SuprimeCam on UT January 5, 2006 as part of a survey for
new Herbig-Haro objects. A five-point dither map with step size of
40'' was used in order to cover the 17'' gaps between the 10
individual CCDs.  A [SII] filter was used with an FWHM of 130 \AA,
central wavelength of 6714 \AA, and peak transmission of 87\%. The
seeing was 0.6'' and the airmass was 1.54.

A second-epoch set of images was obtained at the Gemini-South
telescope during Director's Discretionary Time on March 5 and 7, 2014.
Three broadband filters were used, Sloan {\em r', i',} and {\em z'},
and three narrow-band filters, H$\alpha$, [SII], and a filter with
similar bandwidth in the continuum between the H$\alpha$ and [SII]
filters. Three dithered exposures were obtained in all filters, each
exposure was 40~seconds for the broadband filters, and 200~sec for the
narrow-band filters.  The airmass ranged from 1.26 to 1.48.  The
narrowband images were taken with a seeing of 0.6'' and the broadband
images with 0.7''.

We observed Re50N on 2014 January 19 UT with NASA's Infrared Telescope
Facility located on Mauna Kea, HI.  We used the facility instrument
SpeX (Rayner et al. 1998) to obtain both near-IR images of the region
and low-resolution ($R\approx1200$) spectroscopy of the embedded
source.  Conditions were photometric with seeing roughly $1''$ but
variable as measured by the CFHT seeing monitor.  For imaging, we
obtained seven dithered exposures of Re50 and the A0 V star HD 48481
using each of the MKO J, H, and K-band filters.  Exposure times for
Re50 were 30 sec in J and H-bands, and 5 sec $\times$ 2 coadds in
K-band.  In addition, we obtained five dithered 60 sec exposures of a
nearby empty field in all three filters for sky flats. For the
spectroscopy, we used SpeX in SXD mode with the $0.5''$ slit.  We
nodded the point source in Re50 along the slit in an ABBA pattern for
four individual exposures of 30 sec each, and two of 120 sec each, for
a total exposure time of 6 minutes.  Re50 was observed at an airmass
of $\approx$1.20, with the slit aligned with the parallactic angle to
minimize differential chromatic refraction.  We contemporaneously
observed HD 48481 in SXD mode for telluric calibration.

\section{RESULTS}

In November 2013 we started to follow the Re50 region with the VYSOS
20-inch telescope observing in the $i$-band (Section~2), and within a few
weeks it became obvious that we had caught the object during an
unusual episode of brightening. The light curve is shown in Figure~1.
It indicates that in late 2013 Re50N was brightening at a
rate of $\sim$0.01 mag/day. From the middle of January 2014 and until Orion
disappeared behind the Sun, the rate slowed significantly. When the
region was observed again in September 2014 it was clear that the
brightening had reached a plateau.

We do not know when this brightening began, but we observed the region
in January 2006 with the Subaru telescope and SuprimeCam in search of
Herbig-Haro objects. The image, taken through a [SII] 6717/6731 filter
in 0.6$''$ seeing, is shown in Figure~2a. An identical image from
Gemini-South, also taken in 0.6$"$ seeing, is shown in Figure~2b.

Comparison of the Subaru and Gemini images show that Re50N
brightened dramatically between 2006 and 2014. At
the same time the main object Re50 has faded greatly. Comparison of
the 2014 [SII] image with an identical image taken in the continuum
between [SII] and H$\alpha$ does not reveal the presence of any
shock-excited emission, the light we see is pure continuum, that is,
light reflected from the embedded source.

Figure~3 shows a color mosaic of JHK images, revealing a structured
reflection nebulosity seen mainly in the J filter, and the
illuminating source as a single point source detected only in the
K-filter, suggesting a significant amount of extinction towards the
source. The source corresponds to the previously mentioned VLA~1
source detected in the radio continuum (Morgan et al. 1990, Anglada
1995).  A K-band spectrum of this source is seen in Figure~4,
revealing a featureless very red continuum. This spectral appearance
was already noted by Reipurth \& Aspin (1997) from lower-resolution
spectroscopy.  Such an appearance is in contrast to that of FUors,
which show characteristic deep CO band absorption, thus putting the
FUor classification of this source by Strom \& Strom (1993) in doubt.

Our deep interference filter images reveal several HH objects in the
region.  Such objects are distinguished by their pure emission line
spectra with strong emission in H$\alpha$ and the [SII] 6717/6731
doublet, as well as their characteristic knotty structure.  A compact
group of previously discovered HH knots, HH~65, is located about
3~arcmin northwest of the source (Reipurth \& Graham 1988). It is
unclear if this object is related to HBC~494, since it is located
towards the red lobe of the molecular outflow associated with Re~50,
and should thus be fairly deeply embedded. Moreover, a major molecular
hydrogen flow has been found northwest of Re50 by Stanke et al.
(2002). Two new groups of HH objects are found in our images, here
named HH~1121 and 1122, see Figure~5. HH~1121 consists of two knots,
the western knot is the brighter one, and located at $\alpha_{2000}$
05 40 21.3, $\delta_{2000}$ --07 29 33.  HH~1122 forms a knot complex
just north of Re50 including some knots projected onto Re50 itself; it
was previously detected in near-infrared H$_2$ emission as MHO~161
(Stanke et al. 2002, Davis et al. 2009, 2010). The brightest knot of
HH~1122 is located at $\alpha_{2000}$ 05 40 29.3, $\delta_{2000}$ --07
28 41.  Two faint H$\alpha$ knots are seen just west of Re50N, they
are part of a larger diffuse H$\alpha$ cloud, whose nature is unclear.
Re50N itself is slightly brighter in H$\alpha$ than in the closeby
continuum, suggesting that the illuminating source may have the
H$\alpha$ line in emission. No emission jet could be discerned along
the main axis of Re50N.

\section{DISCUSSION}

The principal question that emerges from the data presented here is,
what is the origin of the dramatic brightening of Re50N?  The two
obvious possibilities are either that the source has undergone a FUor
eruption, or that obscuring material has caused shadows to play on the
surrounding molecular cloud surface. Sometime between 1955 and 1979
the Re50 and Re50N nebulae appeared (Reipurth \& Bally 1986), and this
was ascribed by Strom \& Strom (1986) to a FUor event in the source
based on its P~Cygni profile at H$\alpha$, which they were able to
observe using the reflection nebula as a scattering surface towards
the optically invisible star. Such a FUor event would be able to form
a prominent reflection nebula; a fine example is V1057 Cygni, which
upon its FUor outburst in 1969 formed a large reflection nebula that
expanded as the light from the stellar brightening traveled across the
surrounding cloud (Duncan et al. 1981).

There are several reasons why the FUor interpretation may not hold in
this case. First, the near-infrared spectrum in Figure~4 shows a
smooth red continuum, in contrast to the deep CO bandhead absorption
characteristic of FUors. This is apparently not a temporary spectral
appearance, since Reipurth \& Aspin (1997) saw a similar appearance in
a lower dispersion spectrum obtained in 1996. Second, while it is true
that the H$\alpha$ profile presented by Strom \& Strom (1986) displays
the profound P~Cygni profile characteristic of FUors, such profiles
are sometimes seen in active T~Tauri stars, e.g., AS~353A and
V1331~Cyg (Reipurth et al. 1996, Petrov et al. 2014).  Third, if the
brightening prior to 1979 was due to a FUor outburst, then the present
brightening should be a second FUor outburst subsequent to or on top
of the previous eruption. Such sequential outbursts are not commonly
known in other FUors, in fact only the FUor V1647~Ori has been known
to have had two outbursts (Aspin et al. 2006).

For the above reasons, we suggest that the brightening reported here
has a more straightforward explanation. Reflection nebulae around
young stars are often variable, e.g. the cases of R~Mon (Hubble 1916),
R~CrA (Knox-Shaw 1916), PV~Cep (Cohen et al. 1981), and L483~IRS
(Connelley et al. 2009).  This is ascribed to dusty and clumpy
material orbiting the young stars, causing a play of shadows on a
nearby cloud surface (e.g., Graham \& Phillips 1987).  Although the
appearance of the Re50 region is only sporadically documented, a
fortuitous image taken in 1986 shows Re50N with almost the same
brightness and appearance as today (Reipurth \& Madsen 1989).  In
contrast, Re50 itself used to be a very bright nebula, but it has now
faded greatly to the point of almost disappearing. Comparison of
Figure~1 with the abovementioned earlier images suggest that the
fading started to the west and moved eastwards, as if a curtain of
obscuring material was drawn somewhere along the line of sight to the
illuminating source. These changes occur on longer time scales than
usually seen in reflection nebulae around young stars. This, combined
with the complete absence of any reflection nebula on early
photographic images of the region, suggests that, rather than
resulting from clumpy material orbiting the star, we may be witnessing
the first light escaping from the embedded source as its outflow
cavity begins to break through the surrounding cloud material
(Reipurth \& Bally 1986). Even very slow movement of such obscuring
material can have a significant effect at a distance if it occurs near
the star.

While we favor the shadow play explanation over the FUor hypothesis,
it does not mean that the illuminating source is not heavily
accreting. The strong veiling that is seen in the near-infrared
spectrum is very rare; among the 110 Class~I protostars studied
spectroscopically by Connelley \& Greene (2010), only two have spectra
with similarly strong veiling. In their study, $\sim$85\% of the
sources show spectral features indicative of accretion, with
correlations between veiling, CO emission, and Br$\gamma$ emission. It
is notable that HBC~494 does not show any CO bandhead emission nor
Br$\gamma$ emission, suggesting an abundance of hot dust but little or
no emitting gas. Additionally, the bolometric luminosity of the
source, estimated at $\sim$250~L$_\odot$, is much higher than typical
low-mass young stars, and makes the source one of the most luminous in
the L1641 cloud.  Gramajo et al. (2014) have modelled the energy
distribution of the driving source, and find a substantial disk
accretion rate of $\sim$1.3$\times$10$^{-6}$~M$_\odot$yr$^{-1}$. The
Re50 protostar is evidently in a very active accreting state.

\acknowledgements

We are grateful to Nancy Levenson for allocating Director's
Discretionary Time at Gemini-South for this study (GS-2014A-DD-1), and
to the Gemini-South staff for carrying out the observations.
The Gemini Observatory is operated by the Association of Universities
for Research in Astronomy, Inc., under a cooperative agreement with
the NSF on behalf of the Gemini partnership: the National Science
Foundation (United States), the National Research Council (Canada),
CONICYT (Chile), the Australian Research Council (Australia),
Minist\'erio da Ci\^encia, Tecnologia e Inova\c{c}\~ao (Brazil) and
Ministerio de Ciencia, Tecnología e Innovaci\'on Productiva
(Argentina).
We thank Megan Ansdell and Heather Kaluna for obtaining
early verification images at the UH 2.2m telescope.
This material is based upon work supported by the National Aeronautics
 and Space Administration through the NASA Astrobiology Institute under
 Cooperative Agreement No. NNA09DA77A issued through the Office of Space
 Science.
This research has made use of the SIMBAD database,
 operated at CDS, Strasbourg, France, and of NASA's Astrophysics Data
 System Bibliographic Services.

\clearpage

\begin{figure*}

\plotone{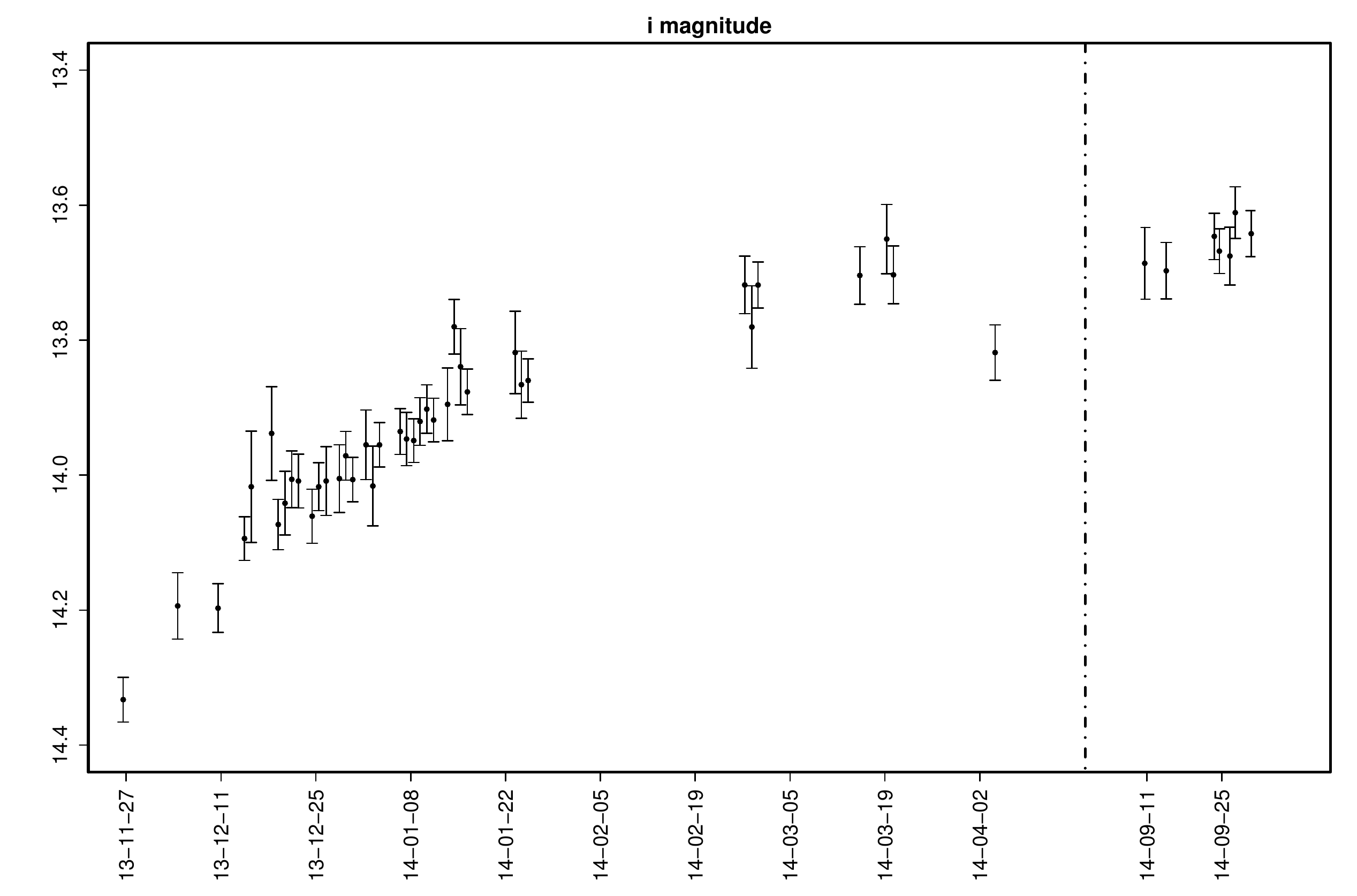}
\caption{A light curve of Re50N obtained in a Sloan {\em i}-filter 
with the VYSOS 20-inch robotic telescope. A 17.8 arcsec aperture was 
used to extract the photometry.  
  \label{fig1}}
\end{figure*}

\begin{figure*}
\plotone{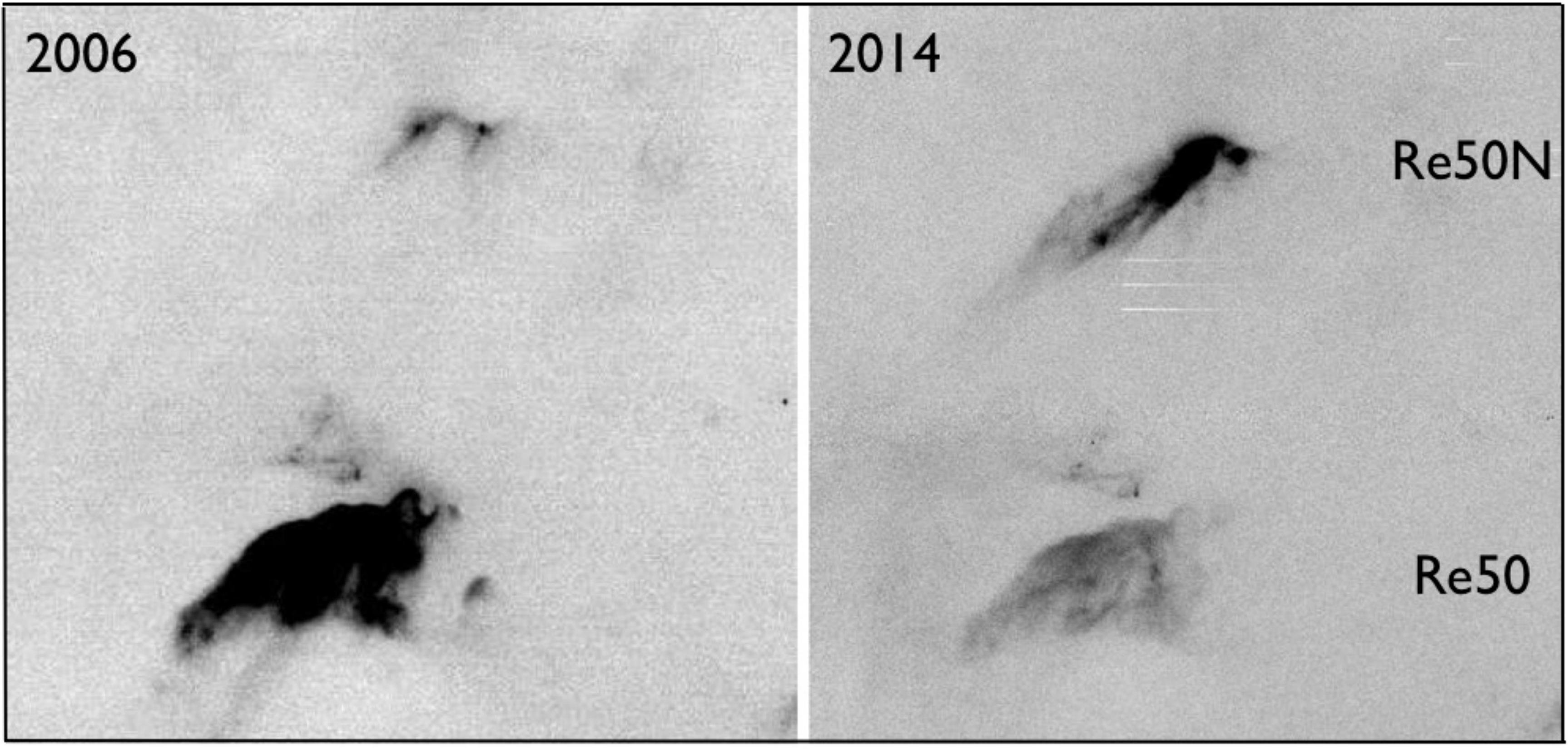}
\caption{The region of Re50 and Re50N observed in 2006 with SuprimeCam 
at the Subaru telescope, and in 2014 with GMOS at the Gemini-S telescope. 
A [SII] filter was used for both images. The seeing was in both cases 
0.6 arcsec. Each image is about 3 arcmin wide. North is up and east is left.
 \label{fig2}}
\end{figure*}

\clearpage

\begin{figure*}
\plotone{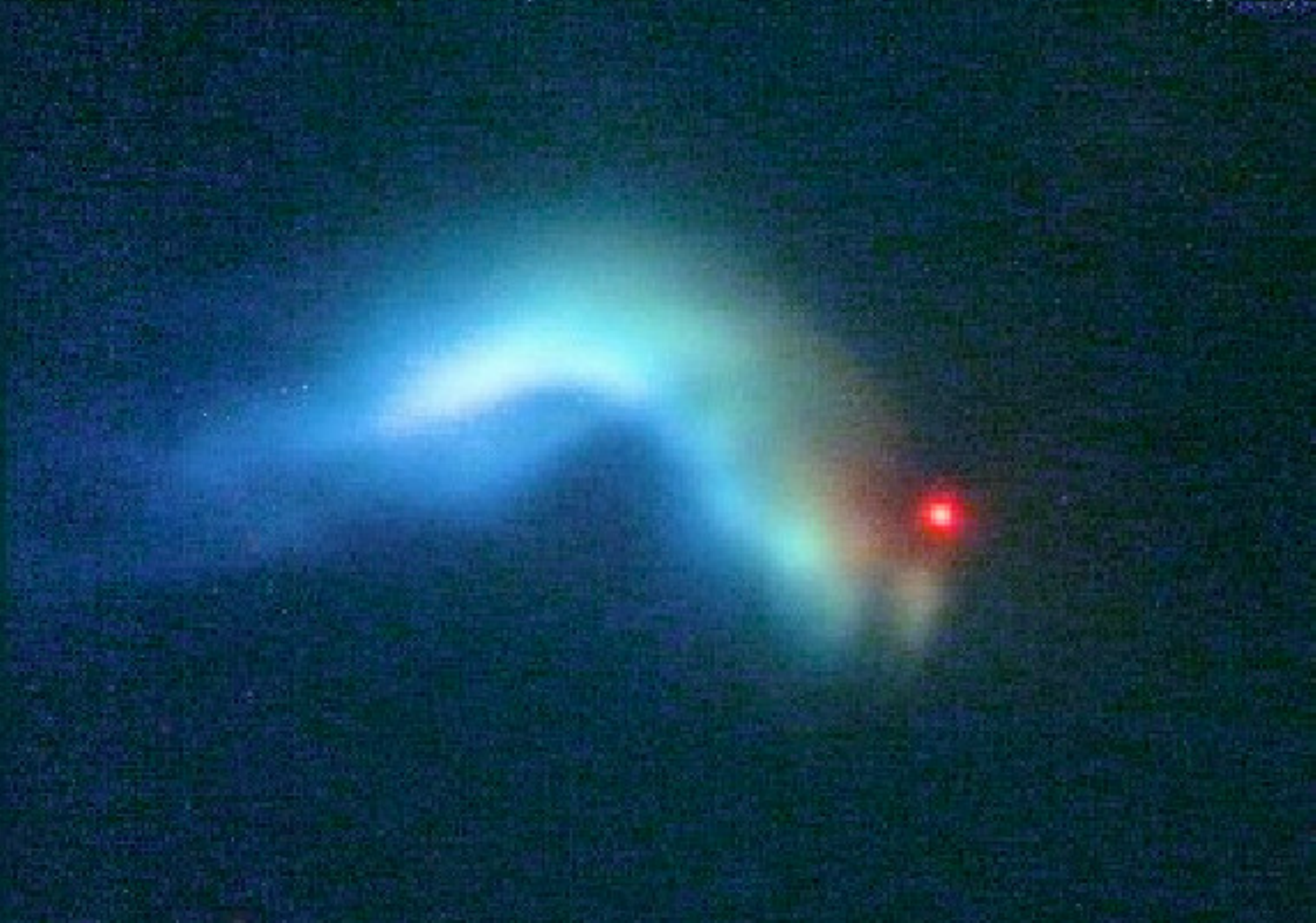}
\caption{A color mosaic of Re50N based on J,H,K images obtained at the IRTF 
(J is blue, H, is green, and K is red). 
The source HBC~494 is the red object, it is detected only in the K-filter.
The image is about 35$"$ wide, with north up and east left.
  \label{fig3}}
\end{figure*}

\vspace{0.3cm}

\begin{figure*}
\plotone{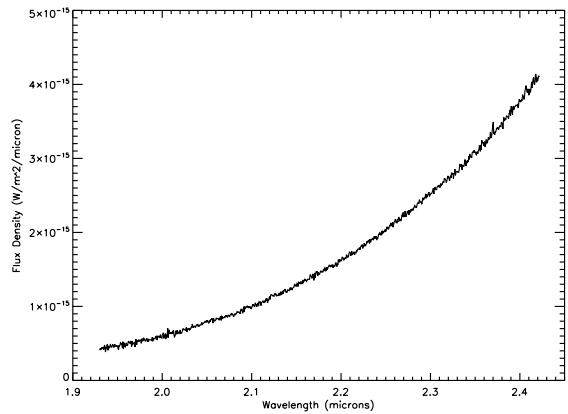}
\caption{A near-infrared spectrum between 1.9 and 2.4~$\mu$m 
of the source HBC~494 obtained with SpeX at the IRTF.
 \label{fig4}}
\end{figure*}

\clearpage

\begin{figure*}
\plotone{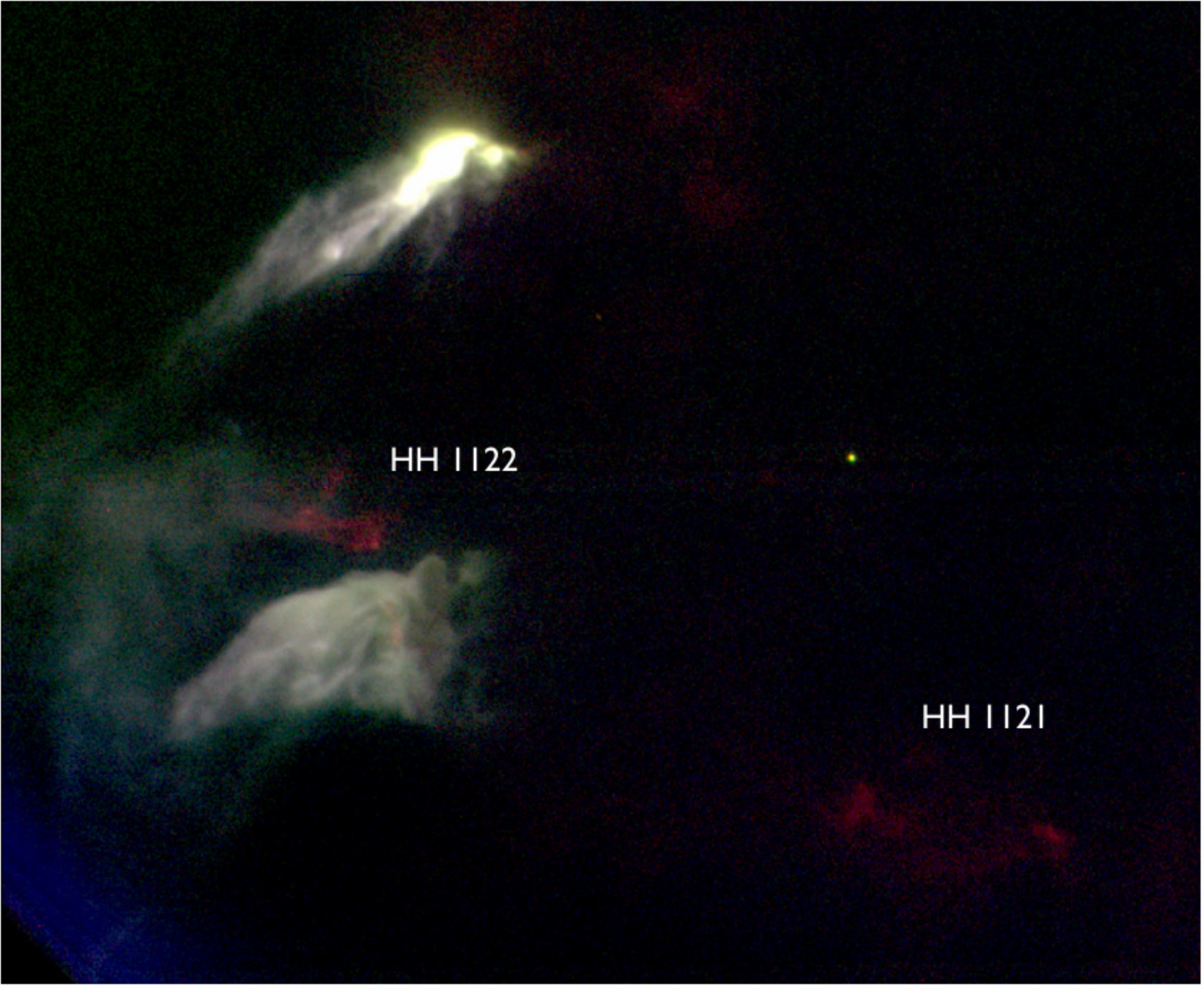}
\caption{Herbig-Haro objects are located around Re50 as seen in this
  color figure where blue is {\em r}, green is {\em i}, and red is the
  sum of {\em z} + H$\alpha$ + [SII] images, all from Gemini-S. 
  The new HH objects HH~1121 and 1122 are seen as faint red features emitting 
  in H$\alpha$ and [SII].  The figure is a little more than 3~arcmin wide. 
  North is up and east is left.  
\label{fig5}}
\end{figure*}

\end{document}